cond-mat/9910141 8 Oct 1999

# On Rational Bubbles and Fat Tails*

## Thomas Lux [1] and Didier Sornette [2,3]

[1] Department of Economics, University of Bonn
Adenauerallee 24 – 42, 53113 Bonn, Germany
lux@iiw.uni-bonn.de
[2] Laboratoire de Physique de la Matière Condensée, CNRS UMR6622
and Université des Sciences, B.P. 70, Parc Valrose; 06108 Nice Cedex 2, France
[3] Institute of Geophysics and Planetary Physics  and Department of Earth and Space Science
3845 Slichter Hall, Box 951567, 595 East Circle Drive
University of California, Los Angeles, California 90095
sornette@cyclop.ess.ucla.edu

*September 1999*

**Abstract**: This paper addresses the statistical properties of time series driven by rational bubbles à la Blanchard and Watson (1982). Using insights on the behavior of multiplicative stochastic processes, we demonstrate that the tails of the unconditional distribution emerging from such bubble processes follow power-laws (exhibit hyperbolic decline). More precisely, we find that rational bubbles predict a 'fat' power tail for both the bubble component and price differences with an exponent $\mu$ smaller than 1. The distribution of returns is dominated by the same power-law over an extended range of large returns. Although power-law tails are a pervasive feature of empirical data, these numerical predictions are in disagreement with the usual empirical estimates. It, therefore, appears that exogenous rational bubbles are hardly reconcilable with some of the stylized facts of financial data at a very elementary level.

JEL-classification: G12, D84, C32

* The idea of this study grew out from discussions at the Workshop 'Facets of Universality: Climate, Biodynamics and Stock Markets' at Giessen University, June 1999. We are grateful to the organizers, Armin Bunde and Hans-Joachim Schellnhuber, for inviting us to this exciting interdisciplinary event. Later versions of the paper benefited particularly from the helpful comments of Holger Drees.

## 1. Introduction

Since the publication of the original contributions on rational expectations (RE) bubbles by Blanchard (1979) and Blanchard and Watson (1982), a huge literature has emerged on theoretical refinements of the original concept and the empirical detectability of RE bubbles in financial data (see Camerer, 1989, and Adam and Szafarz, 1992, for surveys of this literature). Empirical research has largely concentrated on testing for explosive trends in the time series of asset prices and foreign exchange rates. However, it has been shown that tests for explosive roots are quite unreliable in the presence of the more realistic variant of periodically collapsing RE bubbles (Evans, 1991). In fact, as has already been shown by Meese (1986), certain variants of bubble dynamics might look like the outcome of a stationary process. They would, therefore, fool the testing strategy recommended by Hamilton and Whiteman (1985) of comparing the stationarity properties of both asset prices and observable fundamentals. In view of these difficulties, it might be interesting to look for the implications of the bubble models at a more elementary statistical level. In this paper, we derive results on the unconditional distribution of prices, price changes and returns resulting from a standard discrete-time formulation of collapsing bubbles. In this way, we arrive at some neglected observable implications of rational bubbles that can be directly compared with empirical findings. So far, to our knowledge, these elementary statistical properties of RE bubbles have never been investigated. This is astonishing insofar as Blanchard and Watson (1982) themselves refer to the finding of *leptokurtosis* (excessive fourth moments) as a universal characteristic of practically all financial returns. However, their claim that rational bubbles lead to such time series behavior seems to have not been verified in a formal way.[1]

We will, however, not only focus on the kurtosis, as this is a relatively limited measure of deviation from Gaussian statistics, but will refer to a somewhat sharper characterization of the empirical distribution that has emerged from the recent applied literature. In particular, it is now quite universally accepted that the distribution of returns is not only leptokurtotic, but belongs to the class of *fat tailed* distributions. More formally, it has been shown that the tails of the distribution of returns (denoted by $r_t$ in the following) follow approximately a power law:

$$F\left(|r_i| > x\right) \approx c \cdot x^{-\mu} \tag{1}$$

with estimates of the 'tail index' $\mu$ falling in the range 2 to 4 (cf. de Vries, 1994; Pagan, 1996; Guillaume *et al.*, 1997; Gopikrishnan *et al.*, 1998). Introducing somewhat more formal mathematical terminology for later use, the feature expressed in eq. (1) can also be denoted as *regular variation* with index -$\mu$.[2]

---

[1] The argument that leptokurtosis is a signature of speculative bubbles seems widely accepted in the literature, cf. Meese (1986), Camerer (1989).

[2] A function f(x) is said to be regularly varying with index $\alpha$ if $\lim_{x \to \infty} \frac{f(\lambda x)}{f(x)} = \lambda^{\alpha}$, cf. Bingham *et al.* (1987).

As the above findings occur quite uniformly in the returns of all types of financial markets (asset markets and foreign exchange markets as well as future markets and markets for precious metals), the hyperbolic decline in the tails of the empirical distribution appears to be one of the most elementary and pervasive stylized facts for financial markets. It is worth noting that this behavior implies the second and probably the third moment of the distribution are finite, but the fourth and higher moments are infinite. The lack of convergence of the (theoretical) fourth moments not only explains the finding of excessive kurtosis with finite empirical samples but also that its estimation is bound to be unreliable due to inherent fluctuations announcing the theoretical divergence. One may also note that the differences in the tail behavior between the right-hand and left-hand tail have usually found to be not significant which justifies to merge both extremal regions by using absolute values as in eq. (1).

Any satisfactory model of price formation for financial assets should yield time series in accordance with the above stylized fact. Does this hold for the elementary model of RE bubbles? To set the stage, let us first shortly review the basic structure of the RE bubbles approach. Rational expectations require the bubble component in asset prices ($B_t$) to obey:

$$B_t = \delta \cdot E[B_{t+1}], \tag{2}$$

with $\delta$ the discount factor $< 1$. The most elementary variant assumes that the bubble term follows an explosive path with a constant growth rate, i.e. $B_t = a \cdot B_{t-1}$ where the restriction (2) leads to a $= 1/\delta > 1$. The ensuing deterministic dynamics are often considered as appropriate descriptions of price dynamics during periods of hyperinflations, but are less suitable for modeling stock price or foreign exchange movements. In order to overcome the unrealistic picture of ever-increasing deviations from fundamental values, Blanchard and Watson (1982) proposed a model with periodically collapsing bubbles.[3] The bubble component of this model follows an explosive path with probability $\pi$ and collapses to zero with probability $1 - \pi$. In order to allow for the start of new bubble after the collapse, a stochastic component is added to the systematic part of $B_t$:

$$B_t = a_t \cdot B_{t-1} + \varepsilon_t \tag{3a}$$

with: $\qquad a_t = \overline{a} \qquad$ with prob. $\pi$

$\qquad\qquad\qquad a_t = 0 \qquad$ with prob. $1-\pi$, $\tag{3b}$

---

[3] This general formulation allows for both positive and negative bubbles with the sign of each new bubble depending on that of the additive stochastic term in its inception period. However, it is well known that there are conceptual problems with negative bubbles, cf. Diba and Grossman (1987) and we, therefore, deliberately confine ourselves to positive bubbles in the illustrations of our results. As a consequence, in some of the simulations shown below, we did not allow for negative realizations in the period immediately following the crash of a bubble. However, we wish to emphasize that none of our results hinges on this slight modification of the model of eqs. (3), as the resulting downward part of the overall dynamics would share all the qualitative characteristics of the dynamics with positive bubbles only.

and $\varepsilon_t$ IID with mean zero and constant variance $\sigma_\varepsilon^2$. It is easily shown that due to eq. (2) we have: $\bar{a} = (\pi\delta)^{-1} > 1$.

In a recent paper, Fukuta (1998) proposed a refined variant of 'incompletely bursting bubbles' extending eq. (3) to the case of three stochastic states, $a_t \in \{a_1, a_2, a_3\}$, $a_i > 0$, which occur with probabilities $\pi_1$, $\pi_2$, and $\pi_3 = 1 - \pi_1 - \pi_2$. Interestingly, special cases of this still rather simple specification could be shown to serve as discrete-time approximations to certain continuous-time bubbles proposed in the literature.

To make our point in the most general manner, we go one step further and allow for a finite number of stochastic states, i.e. we consider a model with (3b) replaced by (4):

$$a_t = a_i > 0 \quad \text{with probability } \pi_i \tag{4}$$

where $a_i \in \{a_1, a_2, ..., a_n\}$ and $0 \le \pi_1, \pi_2, ..., \pi_n \le 1$, $\sum_{i=1}^{n} \pi_i = 1$.

We also note that all of our results below go through when replacing the discrete distribution of the $a$'s by a continuous distribution function.

The price of the asset (or exchange rate) itself will be composed of its fundamental value and the deviation of it in the form of the bubble term: $p_t = p_{f,t} + B_t$.[4] We make no assumptions on the characteristics of the process driving fundamental values, but rather show below under what conditions the distributional results are dominated by the bubble process rather than the process governing fundamental values. For later reference, the change of the fundamental value in time period t will be denoted by $\eta_t \equiv p_{f,t} - p_{f,t-1}$.

The key towards analysis of the statistical properties of the standard bubble model is the recognition that eq. (3a) describes a particular variant of a *multiplicative stochastic process*. Such processes have been analyzed in some depth by Kesten (1973), Vervaat (1979) and Goldie (1991) and have recently found some interest in physics because of the intermittent character of the ensuing fluctuations (Levy and Solomon, 1996, Sornette and Cont, 1997, Takayasu, Sato and Takayasu, 1997, Sornette, 1998a,b).[5]

Some very general and important results on the statistical behavior of solutions of stochastic difference equations with multiplicative stochastic coefficients are already obtained in Kesten (1975, Proposition 5). In the following, we give the essential results from the papers by Kesten, Vervaat and Goldie which are needed in our ensuing analysis:

---

[4] In the context of foreign exchange markets, one is usually concerned with the *log of the price* (the exchange rate) following the standard formalization of the monetary approach. The bubble term would, then, give the deviation of the log of the actual spot rate from its fundamental solution. We comment on this variant later on. As will be seen, our main conclusions are even somewhat sharper in this case.

[5] The ARCH/GARCH class of time series models is another important economic example of application of multiplicative stochastic processes, cf. de Haan *et al.* (1989)

*Theorem 1* (Kesten, Goldie)

Consider a stochastic difference equation:

$$Y_t = M_t \cdot Y_{t-1} + Q_t, \qquad t = 1,2,\ldots \tag{5}$$

where the pairs { $M_t, Q_t$ } are iid real valued random variables.

(a) If :
   (i)      $E[\ln(|M_t|)] < 0$,

then $Y_t$ converges in distribution and has a unique limiting distribution.

(b) If, additionally, $Q_t/(1-M_t)$ is non-degenerate[6] and there exists some $\mu > 0$ such that:

   (ii)      $0 < E[|Q_t|^\mu] < \infty$,

   (iii)    $E[|M_t|^\mu] = 1$ , and

   (iv)    $E[|M_t|^\mu \cdot \ln^+|M_t|] < \infty$,

then the tails of the limiting distribution are asymptotic to a power law, i.e. they obey a law of the type $\text{Prob}(|Y_t| > x) \approx c \cdot x^{-\mu}$.

The available theory on multiplicative random processes thus assures power-law behavior for a large class of stochastic processes under relatively mild and general conditions. Intuitively, the time-varying multiplicative coefficient $A_t$ yields some sort of 'intermittent' amplification which leads to the heavy power-law tails of the unconditional distribution while the additive noise term $B_t$ preserves the motion from dying out in the course of events.

Note that the existence of a stationary distribution does not hinge on existence of the second or even first moment as μ may assume arbitrary positive values depending on the process under study and only moments of order smaller than μ do exist.

We proceed by considering the behavior of the bubble process itself and the consequences for price changes and returns. We first give results for the Blanchard/Watson model of collapsing bubbles and then proceed to the more general variant in eq. (4).

## 2. Results for the bubble process

### 2.1. Blanchard and Watson model

---

[6] Here, non-degenerate means that $Q_t$ is not a constant time (1 - $M_t$) and the notation $x^+$ denotes max(0, x).

It is straightforward to apply the Kesten theory to the elementary bubble process proposed by Blanchard and Watson. As the multiplicative and additive noise components are given by the pairs $\{a_t, \varepsilon_t\}$, direct application of Theorem 1 under the assumptions of the Blanchard-Watson model yields:

*Result 1:* The Blanchard/Watson bubble process given by eqs (3a,b) has a unique stationary distribution whose tails are asymptotic to a power-law with index:

$$\mu = \frac{\ln(1/\pi)}{\ln(a)} = \frac{\ln(1/\pi)}{\ln(1/\pi) + \ln(1/\delta)} < 1 \quad . \qquad (6)$$

*Proof:* When investigating the stationarity condition (i), $E[\ln(|a_t|)] < 0$, for the case of eqs. (3a,b), one finds that this expectation has a divergence at minus infinity due to the value $a_t = 0$, occurring with probability $1-\pi$, which makes the logarithm diverge. Strictly speaking, $E[\ln(|a_t|)] < 0$ is thus guaranteed. With the assumption of zero mean and constant variance of the additive term, conditions (ii) and (iv) are also fulfilled, and (6) follows immediately from $E[|a_t|^\mu] = 1$.                                                                                                          q.e.d.

Within the Blanchard-Watson model, the bubble terms in asset prices, thus, follow a very heavy-tailed distribution. Although this distribution is unique and stationary, neither its mean nor variance or any higher moment exist. As a consequence, it makes no sense to ask for the *mean deviation* of prices from their fundamental value or the variance of this deviation, as neither of these quantities is finite. In other words, arbitrarily large bubbles will occur with non-negligible probability in this model. It is also worth to add some word on the role of the additive term, $\varepsilon_t$. Note that although we have followed the literature by assuming existence of the second moment of this stochastic variable, our results do not hinge on this assumption. As one can read from condition (ii) of the Kesten theorem, a broad spectrum of distributions including fat-tailed ones would lead to the result of eq. (6). The additive factor is, therefore, of minor importance for the overall characteristics of the distribution of $B_t$.

### 2.2. Generalization with an arbitrary number of $a_i \geq 0$

The general flavor of the above results for the Blanchard/Watson model is, in fact, not restricted to this particular class of speculative bubbles, but rather applies to broad classes of processes. In particular, considering the generalization to an arbitrary number of multiplicative factors in eq. (4), we can show the following:

*Result 2:* For the generalized bubble process which obeys (4) and (2), the following holds: if a unique stationary distribution exists, then its tails are asymptotic to a power-law with index $\mu < 1$.

Hence, all of our qualitative statements above carry over to this more general class, which is also characterized, for example, by non-existence of the expectation of the bubble term.

*Proof:* To prove our second result, we first note that, in this case, a stationary distribution exists if:[7]

$$E[\ln(a_t)] = \sum_{i=1}^{n} \pi_i \ln(a_i) < 0, \tag{7}$$

while the consistency of expectations requires:

$$\sum_{i=1}^{n} \pi_i \cdot a_i = 1/\delta. \tag{8}$$

In the general case, condition (7) may not always be fulfilled. For example, if $a = 1/\delta > 1$ with probability 1 (the explosive case), (7) is clearly violated and no stationary distribution exists. More generally, if the dispersion of the $a_i$ around their mean value $1/\delta$ is small, we will have a permanent increase (on average) of the speculative term, so that the distribution of $B_t$ does not converge. This non-stationarity also implies non-convergence of all positive moments, i.e. $E[|B_t|^\kappa] \rightarrow \infty$ for all $\kappa > 0$. For higher dispersion of the $a_i$, on the other hand, concavity of the log warrants that the stationarity condition (7) will be met.

We concentrate on the latter case and investigate determination of the tail index $\mu$ in the presence of a stationary distribution. The tail index can be computed from:

$$M(\mu) = \sum_{i=1}^{n} \pi_i \cdot a_i^\mu = 1. \tag{9}$$

Taking a closer look at the function $M(\mu)$, we find that the first and second derivative of $M(\mu)$ are given by:

$$M'(\mu) = \sum_{i=1}^{n} \pi_i \cdot a_i^\mu \ln(a_i), \qquad \lim_{\mu \to 0} M'(\mu) = E[\ln(a_i)] < 0, \tag{10}$$

$$M''(\mu) = \sum_{i=1}^{n} \pi_i \cdot a_i^\mu \ln(a_i)^2 > 0. \tag{11}$$

The function $M(\mu)$ is thus convex with $M(0) = 1$. Furthermore, from the consistency requirement (8), we infer that $M(\mu = 1)$ assumes the positive value $1/\delta > 1$.

Taking together these pieces of information, the shape of $M(\mu)$ is as shown in *Fig. 1*. We immediately see that (9) is, therefore, satisfied for some value $\mu_c < 1$. Given that the regularity condition (ii) from Theorem 1 does hold and acknowledging that condition (iv) already follows from eq. (7) and (9) in our case, we, therefore, verify that the solution of (9) for $\mu$ is also less than one in this more general bubble model, for any possible choice of the model parameters. q.e.d.

---

[7] In the notation of the following paragraphs, we dispense with considering absolute values since the multiplicative factors are bound to be non-negative in the rational expectations bubbles anyway.

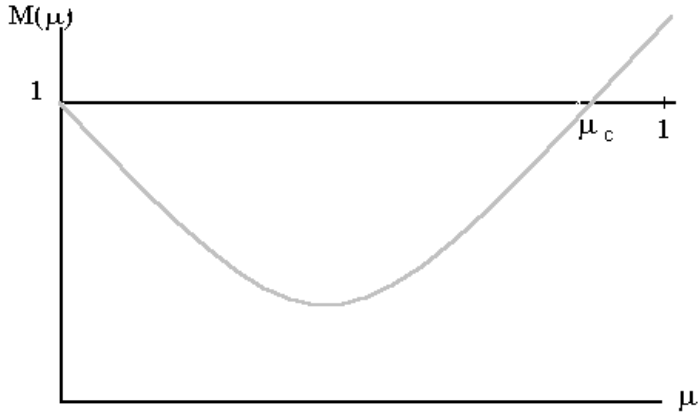

Fig. 1: The Shape of the Function M($\mu$)

The reader may note that it is easy to derive the same result for a continuous distribution of $a_i$'s by replacing the sums in eqs. (7) to (11) by integrals. It is also straightforward to generalize the model by formulating a compound process in which all the $a_i$'s follow continuous distributions and in each time step one of them is chosen with probability $\pi_i$. Hence, for even larger classes of generalized bubble models, the basic restriction (2) assures that the resulting distribution of the bubble component in prices is a particularly fat tailed one with a tail index $\mu < 1$. Again, all the implications that we have drawn from *Result 1* apply here as well. In particular, although the unconditional distribution of prices from the bubble process converges to a stationary distribution, neither the expectation nor the variance of the bubble term exist for all members of these classes of bubble processes.

### 3. Results for price changes

As the next step, we will show that power-law behavior of the bubbles process carries over to the price changes under very mild and general conditions on the dynamics of fundamental values. Denoting price changes by $d_t$, we find that they can be written as

$$d_t = p_{f,t} + B_t - p_{f,t-1} - B_{t-1} = (a_t - 1) \cdot B_{t-1} + \varepsilon_t + \eta_t \qquad (12)$$

where $\eta_t$ is the change of the fundamental value between time t-1 and t. Note that our assumptions on the bubble process imply that $a_t$ and $\varepsilon_t$ are both independent of $B_{t-1}$. Furthermore, since we are dealing with purely exogenously determined bubbles, $\eta_t$ is also independent of all the remaining stochastic variables on the right-hand side of (12). As $d_t$ is determined by multiplicative and additive combinations of four independent random variables, the range of existing moments will be limited by the intersection of the sets of existing moments for the four random variables $a_t$, $B_t$, $\varepsilon_t$, and $\eta_t$. The same applies to its tail behavior which, loosely speaking, will follow that of the 'most fat-tailed' component on the right-hand side. More formally, we can state the following:

> *Result 3:* Assume that asset prices contain a bubble component with a unique stationary distribution whose tails are asymptotic to a power-law with index $\mu < 1$. If, furthermore, $\varepsilon_t$ and $\eta_t$ are both regularly varying and $E[|\varepsilon_t|^\mu]$, $E[|\eta_t|^\mu] < \infty$, then the distribution of price changes is also asymptotic to a power-law with index $\mu$.

*Proof:* For the additive part, our conclusions follow directly from the closure properties of regularly varying functions (cf. Proposition 1.5.7 in Bingham *et al.* (1987)). In particular, if we consider *m* independent random variables possessing regularly varying tails with (real-valued) indices $\alpha_i$, i = 1,..., n, then their sum also has regularly varying tails with index $\alpha_{sum} = \max\{\alpha_1,...,\alpha_n\}$. Hence, provided that the first term on the right-hand side of (12) is regularly varying with index $-\mu$, it suffices to assume regular variation with a smaller index (i.e. larger $\mu$) for the remaining terms. This assumption has already been made for the additive noise term $\varepsilon_t$ and has now been extended to $\eta_t$, in order to assure that the features of the bubble dynamics carry over to the time variation of prices. Two remarks are in order here: first, the assumption of regular variation not only covers the case of power-law tails of $\varepsilon_t$ and $\eta_t$, but also allows for exponential decline of the tail or bounded support, and, hence, includes practically all variants of distributions one is used to in applied work (exponential decline and bounded support can be subsumed under the framework of regular variation by classifying their tail behavior as *rapidly* varying with an index $-\infty$). Second, violation of our condition on $\eta_t$ would only imply that the variation of the fundamental value is even more fat tailed than the bubble term and would, then, dominate the behavior of $d_t$.

It remains to show that the multiplicative part $(a_t - 1) \cdot B_{t-1}$ has the same tail behavior as $B_t$ itself. This follows from results in Breiman (1965)[8] who demonstrates that for two independent random variables $\phi$ and $\chi$ with $\mathrm{Prob}(|\phi| > x) \approx c \cdot x^{-\kappa}$ and $E[\chi^{\kappa+\varepsilon}] < \infty$ for some $\varepsilon > 0$, the random product $\phi \cdot \chi$ obeys $\mathrm{Prob}(|\phi \cdot \chi| > x) \approx E[\chi^\kappa] \mathrm{Prob}(|\phi| > x)$ for $x \to \infty$.

$$\text{q.e.d.}$$

Hence, under the assumption that the $\mu$-th moment of both the distribution of $a_t$ and $\varepsilon_t$ exists

---

(which follows anyway form the conditions for the determination of the tail index $\mu$ for a multiplicative process, cf. Theorem 1), we can state that price differences will be characterized by the same tail behavior provided the distribution of the changes of fundamental value obeys $E[\eta^{\mu+\varepsilon}] < \infty$. Since we know that $\mu < 1$, existence of the mean of $\eta_t$ is sufficient to ensure a tail index equal to $\mu$ of price changes. If, on the other hand, variations of fundamental values would come with a distribution with infinite mean and tail index smaller than $\mu$, the behavior of $\eta_t$ would dominate so that in any case the distribution of price changes in a model with a rational bubble component would be characterized by non-existence of the mean and very heavy tails.

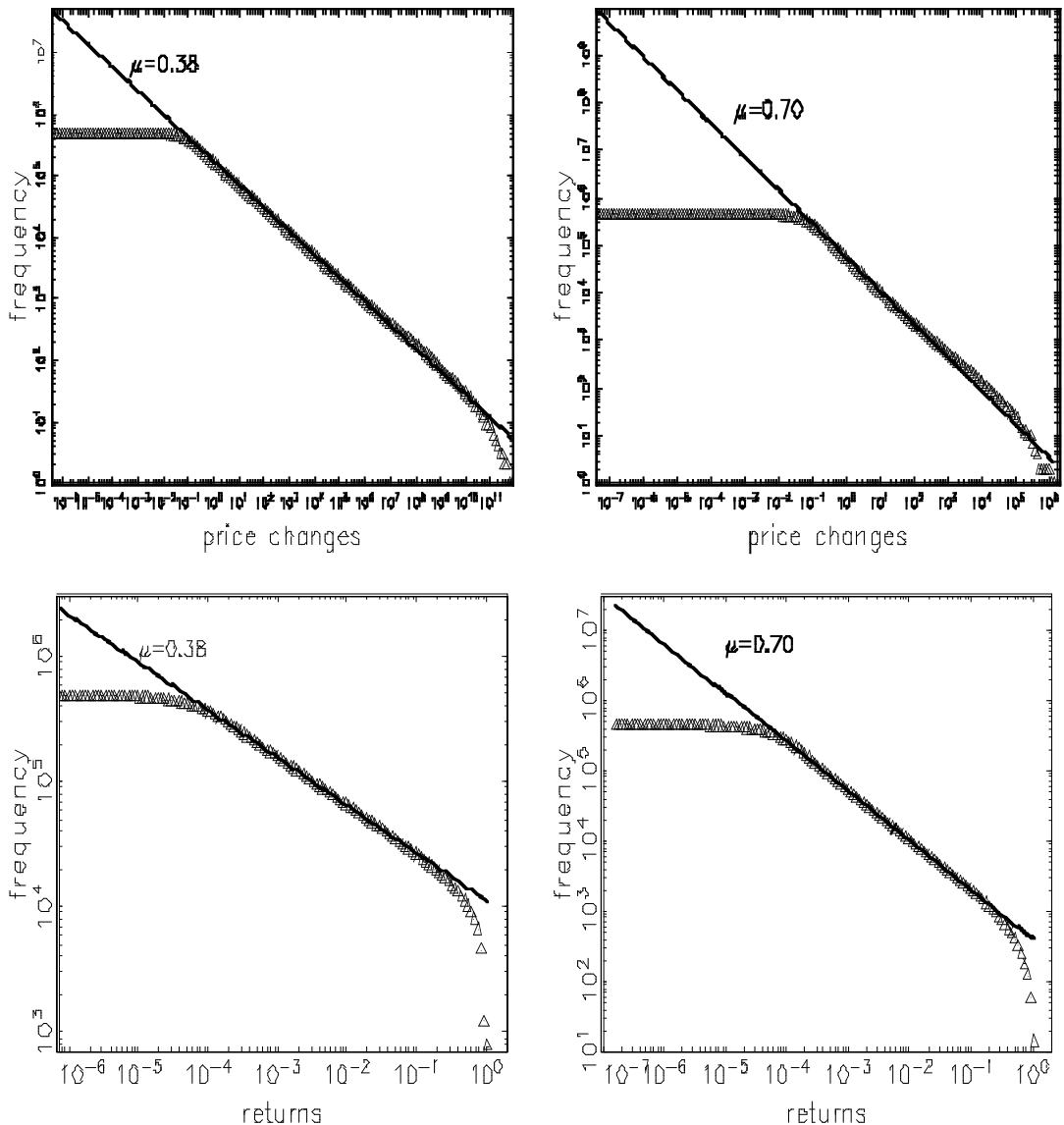

Fig. 2: Frequencies of large price changes and returns for selected bubble processes: the plots

exhibit the complement of the cumulative distribution from simulations over $10^6$ time steps (triangles) and compares them with the theroretically predicted tail behavior (straight line with slope $\mu$). In all cases, the scaling behavior is found to provide a good fit over an extended range of magnitudes. However, while with *price changes*, the entire tail follows a Pareto law with index $\mu$, returns are characterized by deviating behavior at the highest entries.

### 4. Results for returns

Finally, turning to *returns* ($r_t$) which is the usual format of financial time series used in empirical analyses, we note that they can be written as:

$$r_t = \frac{p_{f,t} + B_t - p_{f,t-1} - B_{t-1}}{p_{f,t-1} + B_{t-1}} = \frac{d_t}{p_{f,t-1} + B_{t-1}}, \tag{13}$$

or, alternatively,

$$r_t = \frac{(a_t - 1) \cdot B_{t-1} + \varepsilon_t + \eta_t}{p_{f,t-1} + B_{t-1}}, \tag{14}$$

Although we lack a straightforward theory for the stochastic behavior of this transformation of our original multiplicative model, we can make an 'educated guess' about its likely behavior.

Investigating (13) and (14), we can distinguish two regimes:

(i) for not too large values of the bubble term, $B_t$, the denominator in (13) changes more slowly than the numerator, so that the distribution of returns will be dominated by the variations of the numerator and, hence, will follow approximately the same type of power-law.

(ii) for large bubbles, $B_t \gg p_{f,t}$, the situation changes, however: from (14) we see that when the bubble term increases without bound, the contribution of the fundamental value and its increment to the price becomes negligible. In fact, we find a largest possible return of the order:

$$\max\{|r_t|\} \approx \max\{|\min\{a_i\} - 1|, |\max\{a_i\} - 1|\}. \tag{15}$$

The distribution of returns, will, therefore, follow a power-law with roughly the same exponent $\mu$ as price changes, but with a finite cut-off given in eq. (15). Restricting our attention to positive bubbles, this maximum (absolute) return will always be a negative realization which corresponds to the largest possible crash of the variant under study. For example, in the Blanchard-Watson model of periodically collapsing bubbles, for $a_i < 2$, we have max$\{r_t\}$ -1 ≈ 1.[9]

*Examples: Fig. 2* provides some typical illustrations of the extremal behavior of price changes and returns derived from simulations of various bubble models. The left-hand part shows results from a variant of the original Blanchard-Watson model with parameters $\pi = 0.8$, $a_1 = 1.8$, and $\delta = 1/1.44$. As can be easily computed form (6), the resulting tail index is $\mu \approx 0.38$.

The right-hand panel exhibits a more complicated process in which we allow for compound distributions of the multiplicative factor $a_t$: both $a_1$ and $a_2$ are chosen from a uniform distribution, $a_1 \in [1, 2]$, and $a_2 \in [0, 0.2]$, together with $\pi = 1/1.4$, and $\delta = 1/1.1$. For this variant (which is a special case of the bubble process analyzed in Evans (1991)), one computes $\mu$ from:

$$\frac{1}{1.4}\int_1^2 a_1^\mu da_1 + (1 - \frac{1}{1.4})\int_0^{0.2} 5a_2^\mu da_2 = 1. \tag{16}$$

yielding $\mu \approx 0.7$. In both cases, we have chosen min$\{a_i\} = 0$, simply to avoid numerical overflow in our simulations, and the fundamental value has been kept constant at $p_f = 1000$ for simplicity. As can be seen, our theoretical predictions for the tail index from (6) and (16) are in good agreement with numerical results. Furthermore, the expected cut-off at the maximum return is also well observed.

## 5. Empirical aspects

Neither power-law tails with a tail index $\mu < 1$ nor a finite cut-off are usually found in empirical analyses of the tail behavior of financial data. On the contrary, as stated in the introduction, we have strong evidence of a tail index hovering somewhere between 2 and 4 for practically all relevant empirical records. Right here, the question emerges what kind of results one would have to expect when attempting to 'estimate' the tail index of our simulated price dynamics with rational bubbles. In applied economics literature, a conditional maximum likelihood estimator proposed by Hill has become the standard work tool for estimation of the Pareto exponent of the tails (cf. de Vries, 1994, Pagan, 1996). To compute the Hill tail index estimate, the sample elements of a series (in our case, absolute price differences or returns) are put in descending order: $x_{(n)} \geq x_{(n-1)} \geq ... \geq x_{(n-k)} \geq ... \geq x_{(1)}$ with k the number of observations located in the 'tail' of the distribution. The Hill estimate $\mu_H$ is, then, obtained as:

$$\mu_H = 1 / \frac{1}{k} \sum_{i=1}^{k} [\ln x_{(n-i+1)} - \ln x_{(n-k)}]. \qquad (17)$$

---

[9] Considering assets with negligible intrinsic value ($p_f \approx 0$) we would encounter additional 'regimes'. This happens because small values of $B_t$ in the denominator would lead to large values of returns. The distribution of this type of large returns is governed by the distribution of the smallest absolute values of $B_t$ and, hence, by the central part of the distribution of the noise term, $\varepsilon_t$. For Normally distributed random variables, the distribution of $\varepsilon_t$ raised to the power of minus one follows a power law with exponent $\mu = 1$ (the same holds for a number of alternative distributions). As small denominators may generate much larger returns than the largest crash from the bubble process itself, we would expect a dominating tail behavior with exponent 1 in this case (which can be confirmed by numerical analysis). As neither the underlying scenario nor the resulting time series behavior looks very realistic, we do not follow up this possibility in greater detail.

In order to highlight the results to be expected from a dynamics with rational bubbles, we perform tail index estimation for Monte Carlo simulations of typical time series with a bubble component. We do not expect significant deviations of the Hill estimates from the theoretical results obtained for price differences as we could rigorously prove power-law behavior using the theory of multiplicative stochastic processes and the behavior of the simulated distribution in Fig. 2 nicely confirmed our analytical results. However, we are rather curious for the behavior of the estimates in the case of returns, since the finite cut-off in the latter time series would theoretically imply that we do *not* have a strict power-law at all in this time series but rather exponential decline (rapid variation) at the upper end (and, hence, the estimate $\mu_H$ should tend to infinity). On the other hand, however, we also know that the distribution has an extended scaling region with an approximate power-law behavior.

*Fig. 3* presents typical results for the same time series already investigated in *Fig. 2*. Since the tail size k is not specified in the elementary Hill algorithm, we exhibit results for a bandwidth of tail sizes extending up to a maximum of about 15 per cent of the size of the underlying series. We may first acknowledge that, in fact, there is a very clear correspondence between theory and econometric results for *price differences*. As is to be expected, estimates are somewhat more volatile for very small tail sizes (with only few extreme values used in the estimation), but quickly stabilize around the theoretical tail index. For returns, the picture is, however, not so clear: when varying the number of observations in the tail, k, we see a monotonous decline of the ML estimates. Starting at high numbers indicative of thin tails, the Hill estimates fall quickly towards values smaller than 1 which, then, remain almost constant for an extended range of tail sizes.

As can also be seen, the size of the fundamental value exerts some influence on the exact shape of the curve of the estimates. In particular, with larger $p_f$, the 'attractive' region $\mu < 1$, is more extended which is in agreement with our above arguments: with a higher fundamental value, the denominator in (13) changes more slowly. With changes being dominated by the numerator, the power-law should be visible over a larger range of estimates. Note, that in our simulations, the bubble term raises to orders of $10^{10}$ (cf. *Fig. 2*) while the fundamental value remains constant at a much smaller value. Considering these proportions as unrealistic and using higher fundamental values would only lead to a more extended scaling region in the Hill estimates.[10]

---

[10] Remember, however, that given our theoretical insights, we *expect* unbounded increase of the average size of the bubble term with time, so that in long simulations, bubbles of arbitrary size can occur.

Comparing the simulation results with empirical experience, we see that, in principle, one could select a tail region with 'realistic' estimates between 2 and 4. However, such a result would only hold over a relatively small range of k values. As the empirical Hill estimates do not usually fall below 2 for tail sizes up to 10% of the data, the 'attraction' of the estimates towards values smaller than one (indicative of non-existing means) for relatively moderate tail sizes seems to be at odds with empirical experience.[11]

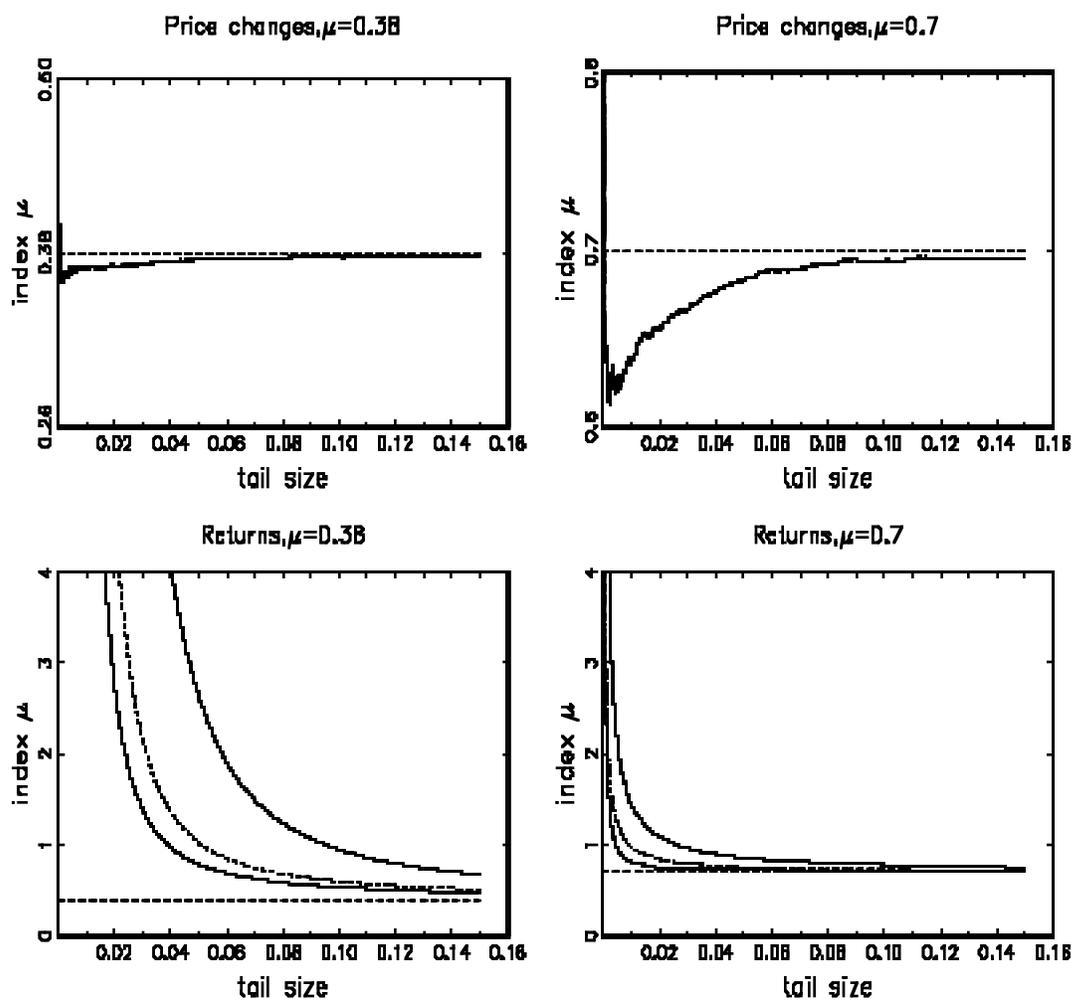

Fig. 3: ML estimates of the tail index with varying tail size. Again results for both price changes and returns are exhibited and the examples are the same as already shown in Fig. 2. For price changes, we see only rather unsystematic deviations from the expected scaling behavior for small tail sizes. For returns, convergence to the theoretically derived tail index

---

[11] The typical appearance of the ML estimates is a hump-shaped curve with a plateau somewhere around 3 which for daily data occurs at a tail size below 5%. Due to increasing bias, the estimates decline for larger k, but do not usually fall below the value 2 with tail sizes as large as 15 to 20% of the data.

(given by the broken line) is also clearly visible, but the finiteness of large returns (due to the cut-off at $r_t \approx 1$) is reflected in a monotonous increase of the estimate at small tail sizes. The speed of convergence is governed by the size of the non-bubble component $p_f$: the three curves in the bottom diagrams are drawn for constant fundamental values $p_f = 100$, $p_f = 500$, and $p_f = 1000$ (from top to bottom).

### 6. Bubbles in foreign exchange markets

So far, we followed the usual practice of the stock market literature in considering speculative bubbles in the *levels* of prices. However, in various models, the fundamental solution is expressed in terms of the *log* of the price and the speculative term is defined as the deviation of the log price from its log fundamental value. This happens quite naturally in models of foreign exchange rates where the market fundamental solution is derived from a monetary macro model. Denoting by $s_t$ the log of the exchange rate, a typical reduced form of (both the fix price and flexible price) monetary macro models leads to (cf., for example, Meese, 1986):

$$s_t = \beta \ E[s_{t+1}] + x_t, \tag{18}$$

with $x_t$ the macro fundamentals, and $\beta \ E[s_{t+1}]$ the expectation term ($0 < \beta < 1$) which enters via uncovered interest parity and the interest elasticity of the demand for money.

With $s_t^*$ denoting the market fundamental solution given the information at time t, rational bubbles come into play with the general solution of (18) without imposition of the transversality condition:

$$s_t = s_t^* + B_t, \quad E[B_{t+1}] = B_t/\beta. \tag{19}$$

Of course, all of our above results still go though with this slight change of the structure of the underlying fundamental model. However, the interesting thing to note is that our previous *Result 3* would now apply to the differences of $s_t$, that is *the log increments of the exchange rate or its returns*. Hence, when dealing with the type of 'monetary approach plus bubbles' model sketched in eqs. (18) and (19, we could skip sec. 5 above and have rigorous analytical results on the statistical behavior of returns readily available from sec. 4. In particular, we were to conclude that this frequently used approach predicts a power-law tail for returns with an index smaller than 1. The consequence of non-existence of both the mean and variance of exchange rate returns, is, however, at odds with empirical findings (cf. de Vries, 1994). Note again, that nothing hinges on the specification of the fundamentals: non-stationarity of $x_t$ could make things only 'worse' (in the sense of smaller $\mu$), but no variant of fundamental dynamics could heal the unrealistic implications for the distribution of returns which are solely derived from the inclusion of a bubble component.

## 7. Conclusion

In the preceding sections, we have pointed out some of the implications for the time series behavior of asset prices being affected by rational bubbles. Our basic findings can be summarized as follows: *the multiplicative nature of the bubble process together with the additive noise guarantees power-.law behavior of the tails and the consistency requirement for the formation of expectations per se, eq. (2), implies non-existence of the mean, variance and all higher moments of the distribution of the bubble component in asset prices. Furthermore, this result can be shown to carry over directly to price changes. For returns, this behavior is reflected in an extending scaling region with similar hyperbolic decline and a sharp fall-off of the distribution thereafter.* When attempting to assess the extremal behavior using standard empirical techniques, the intermediate power-law behavior will also govern the results obtained with returns and, therefore, may point towards a power-law with index smaller than 1. If  bubbles affect the log rather than the level of the price variable, we can rigorously demonstrate such behavior for returns.

The first characteristic, power-law behavior or hyperbolic decline of the outer parts of the distribution function, is, in fact, shared by empirical data from both stock markets and foreign exchange markets. However, the second part which in a sense characterizes the 'degree of fat tailedness' is in contradiction to empirical findings. With a tail index $\mu$ below one, the tails of the distributions from bubble processes are in fact much heavier than those found with empirical data, where one usually computes an index significantly larger than 2 (which assures finiteness of both the mean and variance). The implications of rational bubbles à la Blanchard and Watson are, therefore, hard to reconcile with empirical regularities of financial data at a very elementary level.

Note that our results do not hinge on the frequency of bubble episodes at all. We could chose variants of the model with a high probability of $a_i = 0$ (rare outbreak of bubbles) and, still, all our above conclusions would hold. Probably, we would not encounter such an extended scaling region as in the examples exhibited in *Figs. 2* and *3*, but the power-law tails would show up only for a small part of extreme realizations. Nevertheless, taking into account that extremal analysis has been applied recently to immense data sets of several million observations on the intra-daily level for both stock markets and foreign exchange markets, it seems unlikely that a power-law with index smaller than 1 at the most extreme end of the distribution should have remained unrecognized. However, empirical investigation of high-frequency data (see, for example, Guillaume *et al.*, 1997, and Gopikrishnan *et al.*, 1998) clearly confirmed earlier results obtained with data of daily frequency finding the same scaling in tails as small as 0.01% of the data.

Finally, let us put our present paper into a wider perspective and point out avenues for future research: although we found that rational bubbles in asset prices would lead to statistical behavior at odds with empirics, this, of course, does, not imply that any form of deviations from fundamental valuation can be excluded from an inspection of the statistical characteristics of financial data. In fact, it has been shown recently that certain variants of models with fads and boundedly rational behavior can generate unconditional distributions of returns with the 'right' power-law behavior. For example, Lux and Marchesi (1999) present a micro-simulation model of a large ensemble of boundedly rational speculators in which power-law tails with an index $\mu \approx 3$ (as well as some other stylized facts) emerge endogenously from the trading process. Using similar methods as in the present paper, Aoki (1999) considers a stochastic model of interaction of chartists and fundamentalists. He demonstrates that this framework also gives rise to power-law tails. The tail index of this model depends on the structural parameters of the model, but can be shown to be always larger than 1. Comparing these results with those of the present paper, it, thus, appears that fads models may be better suited than rational bubbles to accommodate the stylized fact of power-law tails with an index in the empirically relevant range.

Of course, the above studies only explore a limited number of possibilities from the large class of fads models, so that it would be premature to draw any general conclusions. We have also not entirely exhausted all types of rational expectations bubbles in this paper: although our analysis covers large classes of exogenous bubbles, there remain some theoretical variants that have not yet been explored. First, we could allow for dependence in the stochastic components $a_t$ and $\varepsilon_t$ as from the economic point of view it may be a sensible generalization to think for example of Markov-switching bubble models where the history determines the likely future development (cf. Johansen, Sornette and Ledoit, 1999, for such a formalization). Available generalizations of Kesten's theory show, that ergodicity of $\{ M_t, Q_t \}$ together with condition (i) in Theorem 1 is sufficient to guarantee existence of a unique stationary distribution (Brandt, 1986). Unfortunately, no results on the tail behavior of this distribution are available so far. Numerical experiments, however, indicate, that power-law behavior with a tail index $\mu < 1$ carries over to this variant. We have also not included the class of fundamental-dependent or intrinsic bubbles in our analysis (Froot and Obstfeld, 1991; Ikeda and Shibata, 1992, 1995). The major problem with this class is that available examples apply certain strong assumptions on the behavior of the fundamental value itself. However, the demonstration by Fukuta (1998) that these models are equivalent to some variants of exogenous bubbles with constant coefficients may open a way to an analysis via the theory of multiplicative processes. This is left for future research.